\documentclass[aps,prd,twocolumn,floatfix,superscriptaddress,balancelastpage,nofootinbib,amsmath,showpacs]{revtex4}
\usepackage{graphicx}
\usepackage{dsfont}
\usepackage{epstopdf}
\usepackage{epsfig}
\usepackage{bm}
\usepackage{dcolumn}
\usepackage{natbib}
\usepackage{amsmath}
\usepackage{amssymb}

\newcommand{\be}{\begin{equation}}
\newcommand{\ee}{\end{equation}}
\newcommand{\bea}{\begin{eqnarray}}
\newcommand{\eea}{\end{eqnarray}}

\def\lsim{\mathrel{\raise.3ex\hbox{$<$\kern-.75em\lower1ex\hbox{$\sim$}}}}
\def\gsim{\mathrel{\raise.3ex\hbox{$>$\kern-.75em\lower1ex\hbox{$\sim$}}}}

\usepackage{slashed}

\newcommand{\mct}{m_{\rm CT}}

\begin{document}

\title{Flavored Dark Matter and the Galactic Center Gamma-Ray Excess}

\author{Prateek Agrawal}
\affiliation{Fermi National Accelerator Laboratory, Theoretical Physics Group, Batavia, IL, 60510}

\author{Brian Batell}
\affiliation{Enrico Fermi Institute, University of Chicago, Chicago, IL,
60637}

\author{Dan Hooper}
\affiliation{Fermi National Accelerator Laboratory, Theoretical Astrophysics Group, Batavia, IL, 60510}
\affiliation{Department of Astronomy and Astrophysics, University of Chicago, Chicago, IL, 60637}

\author{Tongyan Lin}
\affiliation{Kavli Institute for Cosmological Physics,
University of Chicago, Chicago, IL, 60637}

\begin{abstract}
Thermal relic dark matter particles with a mass of 31-40 GeV and that dominantly annihilate to bottom quarks have been shown to provide an excellent description of the excess gamma rays observed from the center of the Milky Way. Flavored dark matter provides a well-motivated framework in which the dark matter can dominantly couple to bottom quarks in a flavor-safe manner. 
We propose a phenomenologically viable model of bottom flavored dark matter that can account for the spectral shape and normalization of the gamma-ray excess while naturally suppressing the elastic scattering cross sections probed by direct detection experiments. This model will be definitively tested with increased exposure at LUX and with data from the upcoming high-energy run of the Large Hadron Collider (LHC). 
\end{abstract}

\pacs{95.35.+d, 95.85.Pw; FERMILAB-PUB-14-069-A-T}

\date{\today}

\maketitle


A robust and wide-ranging experimental effort is currently underway to observe the non-gravitational interactions of dark matter (DM). A major component of this program is focused on the indirect detection of DM through searches for its annihilation products, such as gamma rays, cosmic rays, and neutrinos. Gamma rays from the central region of the Milky Way are particularly interesting in this regard due to the anticipated brightness of the DM annihilation signal and the lack of energy losses or magnetic deflections associated with the high-energy photon signature.

Several independent studies of data from the {\it Fermi} Gamma-Ray Space Telescope have uncovered an excess  
of gamma rays, peaking at $\sim 1-3$ GeV, originating from the direction of the Galactic Center~\cite{Goodenough:2009gk,Hooper:2010mq,Boyarsky:2010dr,Hooper:2011ti,Abazajian:2012pn,Gordon:2013vta,Hooper:2013rwa,Huang:2013pda,Abazajian:2014fta,Daylan:2014rsa}. After being subjected to increasing levels of scrutiny, it appears that this excess cannot be accounted for by any known astrophysical sources or mechanisms (for example, millisecond pulsars~\cite{Hooper:2013nhl}). In terms of energy spectrum, angular distribution, and rate, this signal is remarkably consistent with that long expected from annihilating DM particles (for early predictions, see Ref.~\cite{Bergstrom:1997fj} and references therein). In particular, the recent study of Ref.~\cite{Daylan:2014rsa} concludes that the anomalous gamma-ray emission is well described by a 31-40 GeV DM particle annihilating to $b \bar b$ with a cross section of $\sigma v \simeq (1.7-2.3) \times 10^{-26}\, {\rm cm^3/s}$; in good agreement with that expected for a thermal relic~\cite{Steigman:2012nb}. Here we take this concordance as an indication that the DM may couple dominantly to bottom quarks. While we note that other annihilation modes, such as to light quarks, can provide a good description of the signal's spectral shape, such channels are best fit by annihilation cross sections that are smaller by an order one factor. 

Under the hypothesis that the DM couples preferentially to bottom quarks, we can begin to make inferences about the underlying particle physics theory.  Flavored Dark Matter (FDM) is a framework that naturally leads to flavor-specific DM couplings~\cite{Batell:2011tc,Agrawal:2011ze}. In FDM theories, the DM  particle is part of a flavor multiplet which transforms under the Standard Model (SM) or dark global flavor symmetries. Minimal Flavor Violation (MFV)~\cite{D'Ambrosio:2002ex} can be invoked to suppress new sources of flavor changing neutral currents and simultaneously guarantee the stability of the DM. The stability in this framework is a consequence of Flavor Triality, a $Z_3$ discrete symmetry which is a remnant of the non-Abelian color and quark flavor symmetries~\cite{Batell:2013zwa}. In various guises, FDM has been previously investigated on numerous occasions~\cite{MarchRussell:2009aq,Kile:2011mn,Kamenik:2011nb,Kumar:2013hfa,Lopez-Honorez:2013wla}.

In this work, we propose a model of bottom Flavored Dark Matter ($b$-FDM) and demonstrate that it can account for the Galactic Center gamma-ray excess. The model contains a Dirac fermion transforming as a flavor triplet, of which the third component comprises the cosmological DM. A flavor singlet, color triplet scalar field mediates the interactions between the DM and the Standard Model quarks. An annihilation cross section consistent with the gamma-ray excess can be achieved for perturbative values of the couplings while being consistent with LHC constraints on the colored mediator. For parameters capable of explaining the anomalous gamma-ray signal, the model predicts a direct detection cross section that is consistent with current constraints, but within the near future reach of LUX. The model will be decisively tested with data from the upcoming high-energy run at the LHC. For other investigations motivated by the gamma-ray excess, see Refs.~\cite{Logan:2010nw,Buckley:2010ve,Zhu:2011dz,Marshall:2011mm,Boucenna:2011hy,Buckley:2011mm,Hooper:2012cw,Anchordoqui:2013pta,Buckley:2013sca,Hagiwara:2013qya,Okada:2013bna,Huang:2013apa,Modak:2013jya,Boehm:2014hva,Hardy:2014dea,Finkbeiner:2014sja,Lacroix:2014eea,Hektor:2014kga,Alves:2014yha,simplifiedmodels}.

Throughout this letter,  we will take the DM to be a Dirac fermion and a SM
gauge singlet, with couplings to right-handed down-type quarks. We
take the couplings of $\chi_{b,s,d}$ with quarks to be approximately
flavor diagonal, allowing us to associate each flavor in the dark
sector with a corresponding flavor of quarks. In particular, we take
the lightest of these new particles to be
associated with the $b$-quark, and assume that the heavier flavors
decay into this lightest state.

To allow interactions at the renormalizable level, we must introduce another colored
scalar particle, which we label as $\phi$.\footnote{Although we could have instead considered the case in which the DM candidate is a scalar, the mediator in that scenario is required to be a colored Dirac fermion. As the cross section for the QCD pair production of colored fermions is larger by a factor of $\sim$8 relative to that of colored scalars, this scenario is significantly more restricted by constraints from the LHC.} The relevant terms in the
Lagrangian are given by:
\begin{align}
  \mathcal{L}
  &=
  [m_{\chi}]_{ij} \chi_i \chi^c_j+
{\lambda_{ij}} \chi_{i} d^c_j \; \phi \;  + {\rm h.c.}
\label{quarksingletdown}
\end{align}
where $\chi,\chi^c$ and $d^c$ are 2-component Weyl fermions.

A general flavor structure for either the mass or coupling matrix above would be expected to lead to an unacceptably
large degree of flavor violation. Flavor-safety can be ensured, however, by the MFV ansatz, which postulates that the only source of flavor
violation from new physics are the Yukawa couplings of the SM. In
our case, this can be realized if the DM transforms under one
of the approximate flavor symmetries of the SM: U(3$)_{D}$, U(3$)_{Q}$
or U(3$)_{U}$.  
This restricts the form of the couplings in FDM models. We will consider each case
separately, focusing on those cases in which the DM transforms
under either U(3)$_D$ or U(3)$_Q$.\footnote{We work to leading order in the SM Yukawa couplings. This is justified
as long as the coefficient for higher order corrections does not
overwhelm the small SM Yukawas.}


\begin{figure}
\begin{center}
\includegraphics[width=0.22\textwidth]{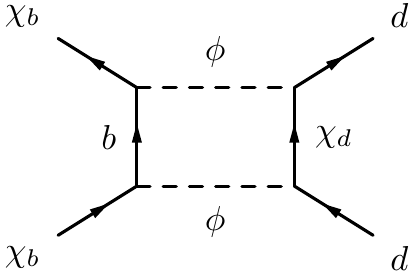}
\qquad
\includegraphics[width=0.22\textwidth]{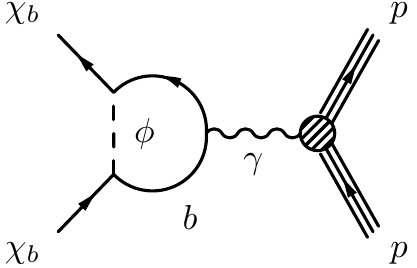}
\caption{\label{fig:dddiag} Feynman diagrams contributing to direct detection.
 }
\end{center}
\end{figure}

\begin{figure}
\begin{center}
\includegraphics[width=0.48\textwidth]{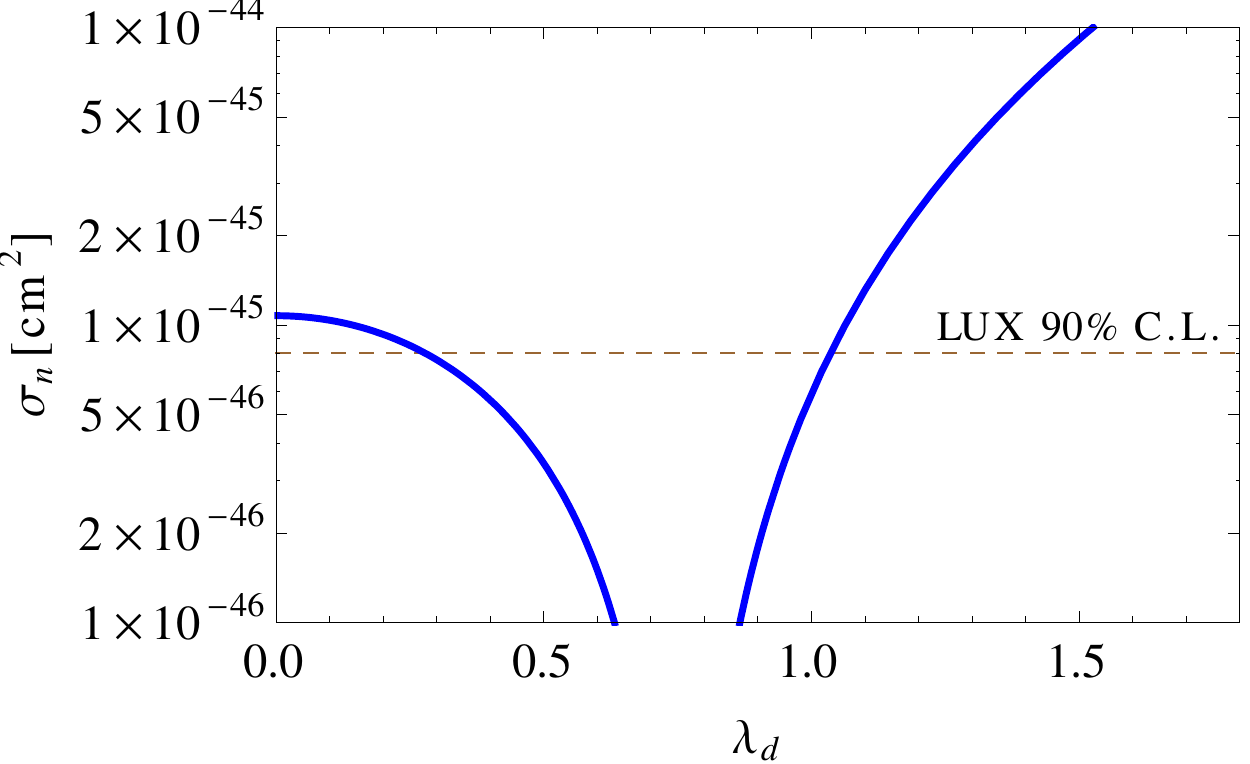}
\caption{\label{fig:ddplot}
The cancellation between the box and photon loop diagrams shown in Fig.~\ref{fig:dddiag}, compared to the (90\% C.L.) upper limits from LUX. Results are shown for $m_{\chi_b}=40$ GeV, $m_{\chi_d}=60$ GeV, $m_{\phi}=725$ GeV, and $\lambda_b=2.16$.}
\end{center}
\end{figure}

Beginning with the case in which the DM transforms under U(3$)_{D}$, the matrix $\lambda$ is constrained to be of the form:
\begin{align}
  \lambda =
  \left( \lambda_0 \mathds{1} +
  \beta  y^{\dagger}_d y_d \right) .
\end{align}
This implies that the couplings of the different quark flavors are approximately universal. The relatively large coupling to first generation quarks that appears in this case leads to large direct detection rates (through the one-loop box diagram shown in Fig.~\ref{fig:dddiag}), potentially in conflict with existing limits. Although it is possible that a cancellation between this box diagram and the one-loop photon exchange diagram (also shown in Fig.~\ref{fig:dddiag}) could allow this constraint to be evaded (see Fig.~\ref{fig:ddplot} and the related discussion)~\cite{QFDM}, it could be argued that direct detection limits disfavor scenarios with universal couplings, and therefore those in which the DM transforms under U(3$)_{D}$. 

In the alternative case, in which the DM transforms under U(3$)_{Q}$, the matrix $\lambda$ is constrained to be proportional to the
down-type Yukawa couplings at leading order:
 \begin{align}
 \label{U3Qlambda}
   \lambda &= \lambda_0 y_d,
 \end{align}
 leading to a hierarchical pattern of couplings. The mass term in this case takes the form:
\begin{align}
 \label{U3Qmass}
  m_{\chi}
  = \left( m_0 \mathds{1} + \Delta m_u \, y_u y^{\dagger}_u +
  \Delta m_d \, y_d y^{\dagger}_d \right).
\end{align}
Since the top Yukawa coupling is large, we generically expect a non-degenerate
spectrum in this case, with the $b$-flavored particle split appreciably from the other two flavor states.

Throughout the remainder of this letter, we will focus on the case in which $\chi$ 
transforms as a triplet under ${\rm U(3)}_Q$, leading to a split spectrum with hierarchical couplings. The Lagrangian of the model is described in Eqs.~(\ref{quarksingletdown},\ref{U3Qlambda},\ref{U3Qmass}) with $m_0$ and 
$\Delta m_u$ of similar magnitude. The mass parameters can be chosen
such that the $b$-flavored component, $\chi_3 \equiv \chi_b$, is the lightest state and thus the DM candidate, while the other states are comparatively heavy and unstable. 

The interaction relevant for most of the cosmology 
and phenomenology in this model is given by:  
\begin{equation}
\label{eq:Lint}
{\cal L} \supset \frac{\lambda_b}{2} \left[  \bar b (1- \gamma_5) \chi_b \phi + \bar \chi_b (1+ \gamma_5) b\, \phi^\dag \right],
\end{equation}
where we have written the fermions as 4-component spinors, and defined  $\lambda_b \equiv \lambda_0 y_b$. The scalar mediator, $\phi$, has the same gauge quantum numbers as the right-handed bottom squark (sbottom) in the supersymmetric (SUSY) version of the SM.  As we will discuss further below, the mass of this particle is constrained by sbottom searches at the LHC to be heavier than about 725 GeV~\cite{sbottom725}. 

\begin{figure}
\begin{center}
\includegraphics[width=0.46\textwidth]{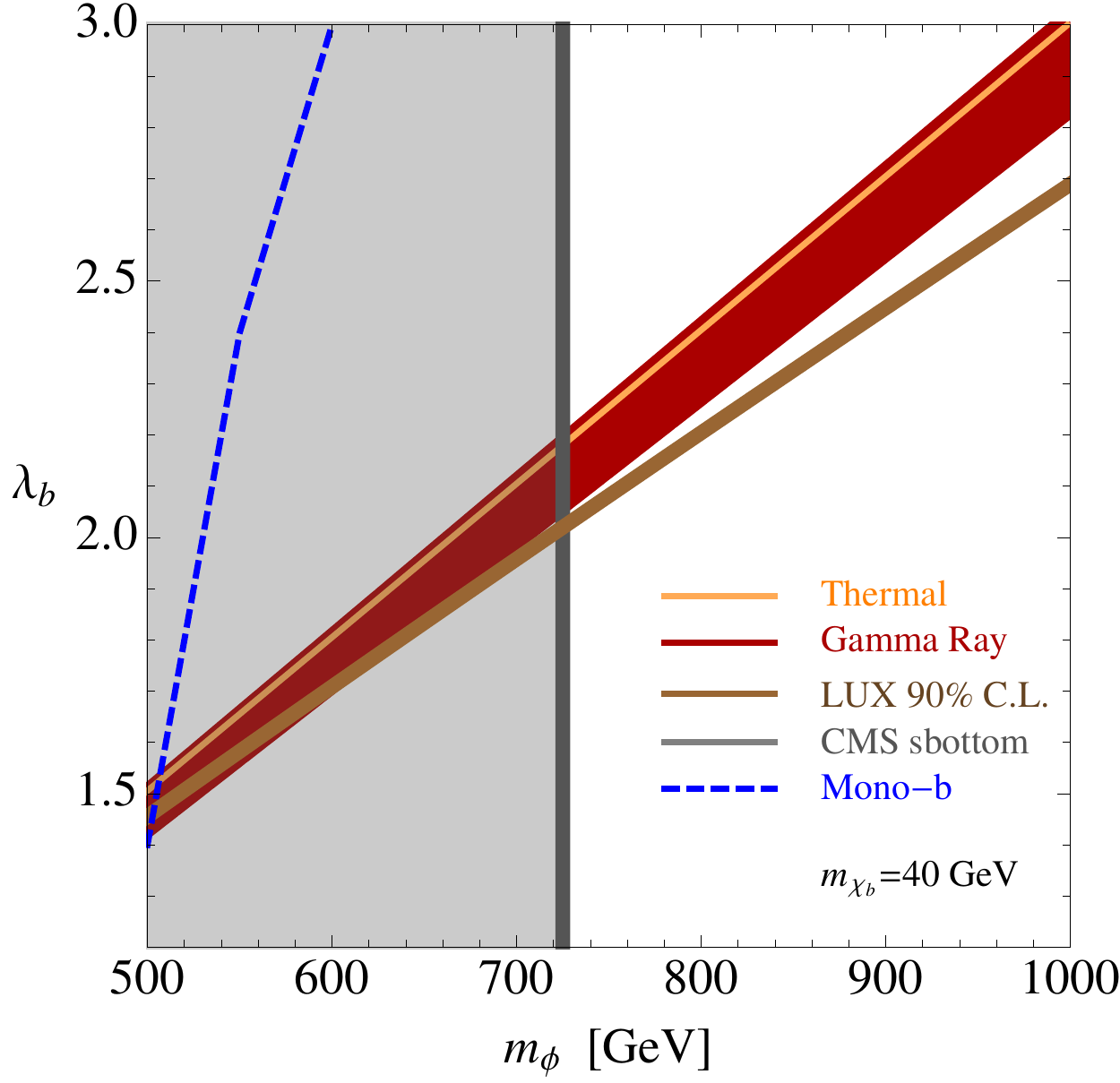}
\caption{The $m_\phi-\lambda_b$ parameter space of the $b$-FDM model for $m_{\chi_b}=40$ GeV. We represent parameters predicting an annihilation cross section consistent with the observed Galactic Center gamma-ray excess (red) and a thermal relic value, $\sigma v = 4.4 \times 10^{-26}\, {\rm cm}^3/{\rm s}$ (orange). We also display the limit reported by the LUX Collaboration (brown). LHC sbottom searches constrain $m_\phi > 725$ GeV (grey). A projection in the mono-$b$ channel with 8 TeV data is also displayed (blue).}
 \label{fig:parameter}
\end{center}
\end{figure}

The thermally averaged annihilation cross section for $\chi_b \bar \chi_b \rightarrow b \bar b$ is given by:
\begin{eqnarray}
\label{eq:sigmav}
\sigma v  & = &  \frac{3 \lambda_b^4 m_{\chi_b}^2 \sqrt{1-m^2_b/m^2_{\chi_b}}}{32 \pi (m_{\chi_b}^2 + m_\phi^2)^2} \times [1+O(v^2)]   \\
 & \approx & 4.4\times 10^{-26} \, {\rm cm}^3/{\rm s}  \left(\frac{\lambda_b}{2.16}\right)^4 
 \left(\frac{m_{\chi_b}}{40\,{\rm GeV}}\right)^2 
 \left(\frac{725\,{\rm GeV}}{m_\phi}\right)^4. \nonumber
\end{eqnarray}
The velocity dependent term is subdominant and has a negligible impact for the model under consideration. Importantly, this equation reveals that a substantial coupling, $\lambda_b \gsim 2$, is required to obtain the observed DM relic abundance (which requires $\sigma v \simeq 4.4 \times 10^{-26}$ cm$^3$/s for a Dirac fermion). The same value of $\lambda_b$ can also accommodate the Galactic Center gamma-ray excess. Such a large coupling can be achieved by taking $\lambda_0$ in Eq.~(\ref{U3Qlambda}) to be large, of order $1/y_b$, or alternatively by taking $\lambda_0 \sim O(1)$ and working in a two Higgs doublet model at large $\tan\beta$. In Fig.~\ref{fig:parameter}, we display parameter regions in the $m_\phi-\lambda_b$ plane where the desired annihilation cross section is obtained. 


There are significant phenomenological consequences of the large couplings required in this model. For hierarchical flavor couplings, as are being focused on here, the dominant contribution to spin-independent scattering arises at
one loop from the charge-charge interaction mediated by photon exchange (see Fig.~\ref{fig:dddiag})~\cite{Agrawal:2011ze}.  Several other
contributions to spin-independent scattering exist and have been previously examined~\cite{Agrawal:2011ze,Kumar:2013hfa,Batell:2013zwa}, but are
subdominant for the model under consideration.

We define the effective DM-nucleon scattering cross section as:
\begin{equation}
\label{eq:sigeff}
\sigma_n \equiv \frac{\mu_n^2 e^2 b^2 Z^2}{\pi A^2},
\end{equation}
where $\mu_n$ is the DM-nucleon reduced mass. For a xenon target (such as that used by LUX),  $Z=54$ and $A=131$. The effective DM-photon coupling, $b$, is given by:
\begin{equation}
b \equiv - \frac{3 Q_b e \lambda_b^2}{64 \pi^2 m_\phi^2} \left[ 1+\tfrac{2}{3} \ln \left( \tfrac{m_b^2}{m_\phi^2}\right)\right],
\end{equation}
where $Q_b = -1/3$. 
For the parameters required to explain the anomalous gamma-ray signal, we find 
\begin{equation}
\sigma_n \approx 1.1 \times 10^{-45}\,{\rm  cm}^2 \times \left( \frac{\lambda_b}{2.16}\right)^4 
 \left( \frac{725\,{\rm GeV}}{m_\phi}\right)^4. 
\end{equation}
This is to be compared to the upper limit reported by the LUX Collaboration, which is $\sigma^{\rm  LUX}_n < 8  \times 10^{-46}\,{\rm cm}^2$ (at the 90$\%$ confidence level) for a 40 GeV DM particle~\cite{Akerib:2013tjd}.
We see that there is a mild degree of tension between the limit reported by LUX and the parameters needed to obtain a thermal relic and explain the gamma-ray excess. There are a number of well-known astrophysics assumptions that enter into direct detection limits, however, and uncertainties associated with the local DM density~\cite{Iocco:2011jz} or velocity distribution~\cite{Vergados:2007nc,Moore:2001vq} could very easily relieve this tension. 
Alternatively, it is worth pointing out that in a 
scenario in which the couplings of the various quark flavors are of
the same order of magnitude, the cancellation between the two diagrams
shown in Fig.~\ref{fig:dddiag} can reduce the elastic scattering
cross section to a value below LUX's current sensitivity. In
Fig.~\ref{fig:ddplot}, we show the effective DM-nucleon cross section,
including the contributions from both of these diagrams, as a function
of the coupling, $\lambda_d$.  For a fairly large range of couplings,
this cancellation significantly reduces the elastic scattering cross
section.

%
%

Collider signals are a powerful test of the $b$-FDM model. The colored mediator, $\phi$, can be directly produced via QCD, 
and its decay $\phi \to b \bar \chi_b$ leads to
final states with $b$-jets and missing transverse energy. This is also the canonical 
signature of the sbottom in SUSY. 
The strongest LHC bound on the $b$-FDM model comes from the sbottom searches described in
Refs.~\cite{sbottom725,Aad:2013ija}, optimized for the pair production of a colored triplet
mediator which decays to a $b$-quark and an invisible DM candidate. CMS results from the 8 TeV run place a limit on the mediator mass of $m_{\phi} > 725$ GeV for a DM mass of 30-40 GeV~\cite{sbottom725}, while the ATLAS sbottom publication gives a slightly weaker bound of $m_{\phi} > 650$ GeV~\cite{Aad:2013ija}. 

These searches targeted a scalar bottom partner in a
simplified SUSY scenario, where QCD pair production of the sbottom is
the dominant mode. In the model under consideration, however, the
coupling $\lambda_b \sim 2 $ is larger than that found in SUSY,
allowing additional processes to significantly contribute to the
production rate. In fact, the cross section (before cuts)
for the production of a single scalar mediator along with a DM
particle, through an off-shell $b$-quark ($gg\to bb^*\to b\phi\chi$)
is slightly larger than that
for QCD mediated $\phi$ pair production. 


To derive limits on such processes, we have simulated the signal for our model in
 {\sc{MadGraph5}} \cite{Alwall:2011uj} with showering and hadronization in
 {\sc{PYTHIA}} \cite{Sjostrand:2006za} and detector simulation with {\sc{DELPHES}}
\cite{deFavereau:2013fsa}. Since the details of CMS analysis are not
currently public, we have applied the event selection as used in the
ATLAS sbottom analysis~\cite{Aad:2013ija} in order to estimate the
impact of the additional processes. In addition to high $p_T$ $b$-jets
and large missing transverse momentum, the event selection includes a
hard cut on the variable $\mct$. Although the additional production
processes in our model have a large rate, the spectrum
associated with these channels is softer since 
only one on-shell heavy mediator leading to a high $p_T$ $b$-jet is
produced, while the $b$-quark
produced through the gluon splitting typically has a smaller $p_T$.
Since the cuts in Ref.~\cite{Aad:2013ija} were optimized for a
much harder processes, we find comparable limits for our
model compared to the sbottom case. The signal rate is only increased
by $\sim$10-20$\%$ in the most sensitive signal region (with $\mct >
350$ GeV), when $\lambda_b$ is taken to be the value required for a
thermal relic.

A complementary signal for this model are events with a single $b$-jet and missing energy. Here, the
production of a single hard $b$-jet along with DM can also be
important, for example $gb \to b \to \phi \chi_b$. This would produce a
mono-$b$ signal, as studied in Ref.~\cite{Lin:2013sca}. 
Our projected 8 TeV limits on $\lambda_b$ as a function of $m_\phi$,
fixing $m_{\chi_b} = 40 $ GeV, are displayed in Fig.~\ref{fig:parameter}. 
The sensitivity of the mono-$b$ channel is
weaker than that for direct $\phi$ pair production; for a thermal relic, only $m_\phi \lesssim 500$ GeV can be tested with the 8 TeV dataset.

We have also performed a collider simulation for the same ATLAS
sbottom search at 14 TeV, including the simulation of the dominant $Z$+jets
background. Even with a conservative estimate, assuming integrated
luminosity of 100 fb$^{-1}$, and applying the same cuts as in the 8 TeV search,
we find an expected reach for $m_{\phi}$ that exceeds 1~TeV.  The mono-$b$ channel will also be able to probe masses up to 700 GeV. Because of the strength of the mono-$b$ signal
and the different kinematics associated with the new production modes
of the mediator, this model could also be distinguished from the SUSY
simplified scenario.  For example, we find that that for $m_\phi \simeq 700$ GeV, 
the mono-$b$ signal is about a factor of 3 larger than in the SUSY case at $\sqrt{s} = 14$ TeV.  

Finally, we note that it appears quite difficult to probe the heavy counterparts to the DM candidate, $\chi_{d,s}$, at the LHC. For hierarchical couplings, $\lambda_i \propto y_i$, the mediator decays  dominantly to $b \bar \chi_b$. Furthermore, while valence $d$ quark initiated processes could, in principle, be relevant for the production of $\chi_d$, the Yukawa suppression in the coupling makes the observation of such events prohibitive. 


The large coupling, $\lambda_b  \sim 2$, required to explain the gamma-ray excess in this model causes the scalar quartic coupling, $\lambda_\phi$, to run rapidly towards negative values as the theory is evolved to higher energies. Around the scale where $\lambda_\phi$ vanishes, the electroweak vacuum is unstable and the theory therefore requires a UV completion. To estimate this scale, we consider the following beta functions, which we obtain from the general results of Ref.~\cite{Luo:2002ti}: 
\begin{eqnarray}
\beta_{\lambda_\phi} & = & 28 \lambda_\phi^2  + 4 \lambda_\phi \lambda_b^2 - 16 \lambda_\phi g_3^2 - 2 \lambda_b^4 + \frac{13}{6}g_3^4, \nonumber \\
\beta_{\lambda_b} & = & 3 \lambda_b^3 - 4 \lambda_b g_3^2, \nonumber  \\
\beta_{g_3} & = & -\frac{41}{6} g_3^3,
\end{eqnarray}
where $16 \pi^2 d\lambda_i / dt = \beta_i $, $t = \log{Q/Q_0}$, and
the coupling, $\lambda_\phi$, is defined through the scalar potential,
$V(\phi) \supset \lambda_\phi (\phi^\dag \phi)^2 $.
  
We have studied the evolution of these couplings from the scale $Q_0 =
m_t$ to higher energies.  In particular, we have fixed $\lambda_b(m_t)
= 2$ and studied several values of $\lambda_\phi(m_t)$, in each case
finding the scale, $\Lambda_{UV}$, where the coupling
$\lambda_\phi(\Lambda_{UV})$ vanishes.  For $\lambda_\phi(m_t) = (1.0,
1.1, 1.2, 1.3, 1.35)$ we find $\Lambda_{UV} \sim (10, 20, 40, 100,
  400)$ TeV. For even larger values of $\lambda_\phi(m_t)$, the
  coupling $\lambda_b$ becomes non-perturbative ($\lambda_b > 4\pi
  /\sqrt{3}$) at a scale below 400 TeV. Thus, we conclude that the
  theory requires additional physics to appear at a scale no higher
  than about this energy. The most straightforward UV completion is a
  SUSY version of $b$-FDM, along the lines of the scenario proposed in
  Ref.~\cite{Batell:2013zwa}. In this case, the superpartner
  contributions to the beta functions tame the UV behavior of the
  theory. For mediator masses above a TeV, relic abundance fixes the
  coupling $\lambda_b > 3$. This requires a UV completion very close
  to the mass of the mediator. Therefore, this simple framework will
  be definitively tested at the next run of the LHC.
 


In summary, as the gamma-ray emission observed from the Galactic
Center has been scrutinized and increasingly well-measured, it has
become only more difficult to explain with known or proposed
astrophysical sources or mechanisms. In contrast, the characteristics
of this gamma-ray excess are in excellent agreement with that
predicted for dark matter in the form of a 31-40 GeV thermal relic,
annihilating to $b\bar{b}$. Using this observation to motivate the
construction of models in which the dark matter preferentially couples
to $b$-quarks, we have discussed a scenario in which the dark matter
is a Dirac fermion that transforms as a flavor triplet, and
annihilates through the $t$-channel exchange of a charged and colored
scalar flavor singlet. This model is flavor-safe and provides a
natural explanation for the stability of the dark matter candidate.
This scenario is also highly predictive and will be definitively
tested in the near future. In particular, this model predicts an
elastic scattering cross section between dark matter and nuclei that
is within the reach of the currently operating LUX experiment.
Additionally, we find that this model will be testable at the upcoming
high-energy run of the LHC over the entire range of viable parameter
space.



\bigskip

{\it Acknowledgements}: 
We would like to thank Matts Buckley and
Reece, and  Raffaele Tito D'Agnolo for helpful discussions.  This work
was supported in part by the Kavli Institute for Cosmological Physics
at the University of Chicago through grant NSF PHY-1125897 and an
endowment from the Kavli Foundation and its founder Fred Kavli.  B.B.
is supported by the NSF under grant PHY-0756966 and the DOE Early
Career Award under grant DE-SC0003930. T.L. thanks the Center for
Future High Energy Physics for their hospitality. 
Fermilab is operated by Fermi Research Alliance, LLC under Contract
No. DE-AC02-07CH11359 with the United States Department of Energy.



\bibliography{bFDM}

\end{document}